\documentclass[12pt]{iopart}
\usepackage{iopams,amssymb}
\newtheorem{theorem}{Theorem}

\newtheorem{definition}{Definition}
\begin{document}

\title[PP-waves with torsion]
{PP-waves with torsion and metric-affine gravity}

\author{Vedad Pasic and Dmitri Vassiliev}

\address{Department of Mathematical Sciences, University of Bath,
Bath BA2 7AY, UK}

\ead{\mailto{V.Pasic@bath.ac.uk} and \mailto{D.Vassiliev@bath.ac.uk}}

\begin{abstract}
A classical pp-wave is a 4-dimensional Lorentzian spacetime
which admits a nonvanishing parallel spinor field;
here the connection is assumed to be Levi-Civita.
We generalise this definition to metric compatible
spacetimes with torsion and describe basic properties of such spacetimes.
We use our generalised pp-waves for constructing
new explicit vacuum solutions of
quadratic metric-affine gravity.
\end{abstract}


\submitto{\CQG}


\section{Introduction}
\label{introduction}

We consider spacetime to be a connected real
4-manifold $M$ equipped with a Lorentzian metric $g$
and an affine connection $\Gamma$.
The 10 independent
components of the (symmetric) metric tensor $g_{\mu\nu}$
and the 64 connection coefficients ${\Gamma^\lambda}_{\mu\nu}$
are the unknowns of our theory.
This approach is known as metric-affine gravity \cite{hehlreview}.

We define our action as
\begin{equation}
\label{action}
S:=\int q(\!R)
\end{equation}
where $q$ is an $\mathrm{O}(1,3)$-invariant
quadratic form on curvature $R\,$.
Independent variation of the metric
$g$ and the connection $\Gamma$ produces Euler--Lagrange
equations which we will write symbolically as
\begin{eqnarray}
\label{eulerlagrangemetric}
\partial S/\partial g&=&0,
\\
\label{eulerlagrangeconnection}
\partial S/\partial\Gamma&=&0.
\end{eqnarray}
Our objective is the study of the
combined system of field equations
(\ref{eulerlagrangemetric}), (\ref{eulerlagrangeconnection}).
This is a system of $10+64$
real nonlinear partial differential equations
with $10+64$ real unknowns.

The motivation for choosing a model of gravity which is purely quadratic
in curvature is explained in Section 1 of \cite{annalen}.
Basically, we are hoping to describe
physical phenomena whose characteristic wavelength is sufficiently small
and curvature sufficiently large. Also, the choice of action
which is homogeneous (of degree 2) in curvature means that we
are looking for vacuum solutions.

The Yang--Mills action
for the affine connection is a special case of (\ref{action}) with
\begin{equation}
\label{YMq}
q(\!R)=q_{\mathrm{YM}}(\!R):=
R^\kappa{}_{\lambda\mu\nu}\,
R^\lambda{}_\kappa{}^{\mu\nu}\,.
\end{equation}
With this choice of $q$ equation (\ref{eulerlagrangeconnection}) is
the Yang--Mills equation for the affine connection.

The quadratic form $q$ appearing in (\ref{action}) is a generalisation of
(\ref{YMq}). The general formula for $q$ contains 16 different $R^2$-terms
with 16 coupling constants. This formula is given in Appendix B
of \cite{annalen}. An equivalent formula can be found in
\cite{Esser,hehlandmaciasexactsolutions2}.

\begin{definition}
\label{definition of riemannian}
We call a spacetime $\{M,g,\Gamma\}$
\emph{Riemannian}
if the connection is Levi-Civita
(i.e.
${\Gamma^\lambda}_{\mu\nu}=\left\{{{\lambda}\atop{\mu\nu}}\right\}$),
and \emph{non-Riemannian}
otherwise.
\end{definition}

The aim of this paper is to find new non-Riemannian
solutions of the field equations
(\ref{eulerlagrangemetric}), (\ref{eulerlagrangeconnection}).
These new solutions will be constructed explicitly
and the construction will turn out to be very similar
to the classical construction of a pp-wave, only with torsion.
In fact, the generalisation of the concept of a pp-wave
to spacetimes with torsion is the main tool in our analysis
and a useful spin-off which might be of wider differential
geometric interest.

The paper has the following structure.
In Section \ref{Classical pp-waves} we recall basic facts concerning
classical pp-waves (without torsion).
In Section \ref{PP-waves with torsion} we define the notion of
a generalised pp-wave (with torsion) and list the main properties of such spacetimes.
In Section \ref{Explicit representation of our field equations}
we write down explicitly our field equations
(\ref{eulerlagrangemetric}), (\ref{eulerlagrangeconnection})
and in Section \ref{PP-wave type solutions of our field equations}
we present pp-wave solutions of these field equations.
Theorem \ref{main theorem}
of Section \ref{PP-wave type solutions of our field equations}
is the main result of our paper.
We discuss our results in Section \ref{Discussion}.
Finally, \ref{Spinor formalism for generalised pp-spaces}
and
\ref{Derivation of the second field equation} contain some auxiliary
mathematical facts.

\section{Notation}
\label{Notation}

Our notation follows \cite{King and Vassiliev,pseudo,annalen}. In particular,
we denote local coordinates by $x^\mu$, $\mu=0,1,2,3$,
and write $\partial_\mu:=\partial/\partial x^\mu$.
We define the covariant derivative of a vector field as
$\nabla_{\mu}v^\lambda:=\partial_\mu v^\lambda
+{\Gamma^\lambda}_{\mu\nu}v^\nu$
and
torsion as
${T^\lambda}_{\mu\nu}:=
{\Gamma^\lambda}_{\mu\nu}-{\Gamma^\lambda}_{\nu\mu}\,$.
We say that our connection $\Gamma$ is metric compatible
if $\nabla g\equiv0$.
The Christoffel symbol is
$\left\{{{\lambda}\atop{\mu\nu}}\right\}:=
\frac12g^{\lambda\kappa}
(\partial_\mu g_{\nu\kappa}
+\partial_\nu g_{\mu\kappa}
-\partial_\kappa g_{\mu\nu})$.
The interval is
$\rmd s^2:=g_{\mu\nu}\,d x^\mu\,d x^\nu$.

We define curvature as
${R^\kappa}_{\lambda\mu\nu}:=
\partial_\mu{\Gamma^\kappa}_{\nu\lambda}
-\partial_\nu{\Gamma^\kappa}_{\mu\lambda}
+{\Gamma^\kappa}_{\mu\eta}{\Gamma^\eta}_{\nu\lambda}
-{\Gamma^\kappa}_{\nu\eta}{\Gamma^\eta}_{\mu\lambda}\,$,
Ricci curvature as
$Ric_{\lambda\nu}:={R^\kappa}_{\lambda\kappa\nu}\,$,
scalar curvature as $\mathcal{R}:=Ric^\lambda{}_\lambda\,$,
and trace-free Ricci curvature as
$\mathcal{R}ic:=
Ric-\frac14\mathcal{R}g$.
We denote Weyl curvature by $\mathcal{W}$;
here, as in \cite{pseudo,annalen}, Weyl curvature is understood
as the irreducible piece of curvature defined by conditions
(\ref{symmetries of Riemannian curvature 1}), (\ref{symmetries of Riemannian curvature 2})
and $Ric=0$.

We employ the standard convention of raising and lowering tensor indices
by means of the metric tensor.
Some care is, however, required when
performing covariant differentiation:
the operations of raising and lowering of indices do not
commute with the operation of covariant differentiation
unless the connection is metric compatible.

Given a scalar function $f:M\to\mathbb{R}$ we write for brevity
\[
\int f:=\int_Mf\,\sqrt{|\det g|}\,\rmd x^0\rmd x^1\rmd x^2\rmd x^3\,,
\qquad\det g:=\det(g_{\mu\nu})\,.
\]

We define the action of the Hodge star on a rank $q$ antisymmetric tensor as
\[
(*Q)_{\mu_{q+1}\ldots\mu_4}\!:=(q!)^{-1}\,\sqrt{|\det g|}\,
Q^{\mu_1\ldots\mu_q}\varepsilon_{\mu_1\ldots\mu_4}
\]
where $\varepsilon$ is the totally antisymmetric quantity,
$\varepsilon_{0123}:=+1$.
When we apply the Hodge star to curvature we have a choice between
acting either on the first
or the second pair of indices, so we introduce two
different Hodge stars: the left Hodge star
\[
({}^*\!R)_{\kappa\lambda\mu\nu}:=
\frac12\,\sqrt{|\det g|}
\ R^{\kappa'\lambda'}{}_{\mu\nu}\,
\varepsilon_{\kappa'\lambda'\kappa\lambda}
\]
and the right Hodge star
\[
(R^*)_{\kappa\lambda\mu\nu}:=
\frac12\,\sqrt{|\det g|}
\ R_{\kappa\lambda}{}^{\mu'\nu'}\,
\varepsilon_{\mu'\nu'\mu\nu}\,.
\]
Note that in the general metric-affine setting
curvature is not necessarily antisymmetric in the first pair of indices so
use of the left Hodge star really makes sense
only in metric compatible spacetimes.

We use the term ``parallel''
to describe the situation when the covariant derivative of some spinor
or tensor field is identically zero.

We do not assume that our spacetime admits a (global) spin structure,
cf. Section 11.6 of \cite{Nakahara}.
In fact, our only topological assumption is connectedness.
This does not prevent us from defining and parallel transporting
spinors or tensors locally.

\section{Classical pp-waves}
\label{Classical pp-waves}

In this section spacetime is assumed to be Riemannian, see Definition
\ref{definition of riemannian}.

\begin{definition}
\label{definition 1 of a pp-space}
A \,\emph{pp-space} is a Riemannian spacetime which
admits a nonvanishing parallel spinor field.
\end{definition}

We will call the metric of a pp-space \emph{metric of pp-wave} or
simply \emph{pp-metric}.
Such metrics were introduced by Peres \cite{peres,peresweb}
who used the equivalent
Definition \ref{definition 2 of a pp-space}
given further on in this section.

Throughout this paper we denote the nonvanishing parallel spinor field
by $\chi=\chi^a$ and assume that this spinor field is \emph{fixed}.
Note that
\begin{itemize}
\item
a nonvanishing parallel spinor can be scaled by a nonzero complex factor
(there is no natural normalisation), and
\item
in flat space there are two linearly independent
nonvanishing parallel spinor fields.
\end{itemize}
Fixation of the spinor field $\chi$ allows us to avoid ambiguity in subsequent arguments.

Put
\begin{equation}
\label{formula for l}
l^\alpha:=\sigma^\alpha{}_{a\dot b}\,\chi^a\bar\chi^{\dot b}
\end{equation}
where the $\sigma^\alpha$ are Pauli matrices, see
\ref{Spinor formalism for generalised pp-spaces}
for notation.
Then $l$ is a nonvanishing parallel real null vector field.
Define also the real scalar function
\begin{equation}
\label{phase}
\varphi:M\to\mathbb{R},\qquad
\varphi(x):=\int l\cdot\rmd x\,.
\end{equation}
This function is called the \emph{phase}.
It is defined uniquely up to the addition
of a constant and possible multivaluedness
resulting from a nontrivial topology of the manifold.

The 3-manifolds $\tilde M=\{\varphi=\mathrm{const}\}$ are called
\emph{wave fronts}. Let us fix a particular wave front $\tilde M$,
take a pair of points $\tilde p,\tilde q\in\tilde M$, and a curve
$\tilde\gamma\subset\tilde M$ connecting these points. Take a
4-vector tangent to $\tilde M$ at $\tilde p$ and parallel transport
it in accordance with the Levi-Civita connection along
$\tilde\gamma$. It is easy to see that the resulting 4-vector will
be tangent to $\tilde M$ at $\tilde q$. This means that the
Levi-Civita connection $\Gamma$ over $TM$ admits a natural
restriction to a connection $\tilde\Gamma$ over $T\tilde M$. (The
latter cannot be interpreted as the Levi-Civita connection
corresponding to the restriction of our Lorentzian 4-metric to the
3-manifold $\tilde M$ as this restricted metric is degenerate.) An
important property of pp-spaces is that the connection
$\tilde\Gamma$ is flat. This is why pp-spaces are often called
``plane-fronted gravitational waves with parallel rays''.

The fact that the wave fronts are flat motivates the following definitions.

\begin{definition}
\label{definition of a transversal vector field}
We say that a complex vector field $u$ is \emph{transversal}
if $l_\alpha u^\alpha=0$.
\end{definition}

\begin{definition}
\label{definition of a plane wave}
We say that a complex vector field $v$ is a \emph{plane wave} if
$u^\alpha\nabla_\alpha v^\beta=0$ for any transversal vector field $u$.
\end{definition}

Of course, $l$ itself is transversal and a plane wave.

Put
\begin{equation}
\label{formula for F}
F_{\alpha\beta}:=\sigma_{\alpha\beta ab}\,\chi^a\chi^b
\end{equation}
where the
$\sigma_{\alpha\beta}$ are ``second order Pauli matrices''
(\ref{second order Pauli matrices}). Then $F$ is a nonvanishing
parallel complex 2-form with the additional properties
$*F=\pm\rmi F$ and $\det F=0$. It can be written as
\begin{equation}
\label{decomposition of F}
F=l\wedge a
\end{equation}
where $a$ is a complex vector field satisfying
$a_\alpha a^\alpha=l_\alpha a^\alpha=0$, $a_\alpha\bar{a}^\alpha=-2$.
The vector field~$a$
is defined uniquely up to the addition of
\[
\{\mathrm{arbitrary\ complex\ valued\ scalar\ function}\}\times l\ .
\]
We can impose an additional restriction on our choice of
$a$ requiring that $a$ be a plane wave.
Under this restriction the vector field $a$
is defined uniquely up to the addition of
\[
\{\mathrm{arbitrary\ complex\ valued\ scalar\ function\ of\ }\varphi\}
\times l
\]
and
\begin{equation}
\label{covariant derivative of a}
\nabla_\alpha a_\beta=
p\,l_\alpha l_\beta.
\end{equation}
where $p:M\to\mathbb{C}$ is some scalar function.

Throughout this paper our choice of the vector field $a$ is assumed to be fixed.
This implies, in particular, that the function $p$ appearing in
(\ref{covariant derivative of a}) is fixed.

It is known,
see Section 4 in \cite{Alekseevsky} or Section 3.2.2 in \cite{Bryant},
that Definition \ref{definition 1 of a pp-space}
is equivalent to the following

\begin{definition}
\label{definition 2 of a pp-space}
A \,\emph{pp-space} is a Riemannian spacetime whose metric can be written
locally in the form
\begin{equation}
\label{metric of a pp-wave}
\rmd s^2=
\,2\,\rmd x^0\,\rmd x^3-(\rmd x^1)^2-(\rmd x^2)^2 +f(x^1,x^2,x^3)\,(\rmd x^3)^2
\end{equation}
in some local coordinates $(x^0,x^1,x^2,x^3)$.
\end{definition}

We do all our practical calculations in coordinates
(\ref{metric of a pp-wave}) and with Pauli matrices
(\ref{Pauli matrices for pp-metric}).
Of course, the choice of local coordinates in which the pp-metric
assumes the form (\ref{metric of a pp-wave}) is not unique.
We will restrict our choice to those coordinates in which
\begin{equation}
\label{explicit l and a}
\chi^a=(1,0),
\qquad
l^\mu=(1,0,0,0),
\qquad
a^\mu=(0,1,\mp\rmi,0).
\end{equation}
With such a choice formula (\ref{phase}) reads
$\varphi(x)=x^3+\mathrm{const}$.

The remarkable property of the metric
(\ref{metric of a pp-wave}) is that the corresponding
curvature tensor $R$ is linear in $f$:
\begin{equation}
\label{curvature of a pp-space special}
R_{\alpha\beta\gamma\delta}=
-\frac12(l\wedge\partial)_{\alpha\beta}\,(l\wedge\partial)_{\gamma\delta}f
\end{equation}
where
$(l\wedge\partial)_{\alpha\beta}:=l_\alpha\partial_\beta-\partial_\alpha l_\beta$.
Simplicity of the formula for curvature
was the main motivation for Peres
when he introduced \cite{peres,peresweb} the concept of a pp-space (pp-wave).

Observe now that in our special local coordinates $f$ satisfies the equations
\begin{equation}
\label{potential equation for f}
l^\mu\partial_\mu f=0,\qquad a^\mu\partial_\mu f=p/2
\end{equation}
where $p$ is the function from (\ref{covariant derivative of a}).
Equations (\ref{potential equation for f}) are invariantly defined
equations for a scalar function $f:M\to\mathbb{R}$. These equations
allow us to give an invariant interpretation of our function $f$
as a \emph{potential} for a pp-space.
Equations (\ref{potential equation for f}) specify the gradient of~$f$
along wave fronts, and, consequently, they define $f$ uniquely
up to the addition of an arbitrary real valued scalar function of $\varphi$.

Formula (\ref{curvature of a pp-space special}) can now be rewritten
in invariant form
\begin{equation}
\label{curvature of a pp-space}
R=-\frac12(l\wedge\nabla)\otimes(l\wedge\nabla)f
\end{equation}
where $l\wedge\nabla:=l\otimes\nabla-\nabla\otimes l$.
Indeed,
in our special local coordinates all the terms with connection
coefficients in the RHS of (\ref{curvature of a pp-space}) cancel out and
(\ref{curvature of a pp-space}) turns into (\ref{curvature of a pp-space special}).
As both sides of (\ref{curvature of a pp-space}) are tensors
formula (\ref{curvature of a pp-space}) holds in any coordinate system.

It is easy to see that the curvature of a pp-space has only two
irreducible pieces, trace-free Ricci and Weyl. Ricci curvature is
proportional to $l\otimes l$
whereas Weyl curvature is a linear combination of
$\mathrm{Re}\left((l\wedge a)\otimes(l\wedge a)\right)$
and
$\mathrm{Im}\left((l\wedge a)\otimes(l\wedge a)\right)$.

\section{PP-waves with torsion}
\label{PP-waves with torsion}

The most natural way of generalising the concept of a classical pp-space
is simply
to extend Definition \ref{definition 1 of a pp-space} to general metric
compatible spacetimes, i.e. spacetimes whose connection is not necessarily
Levi-Civita. However, this gives a class of spacetimes which is too wide
and difficult to work with. We choose to extend the classical definition in
a more special way better suited to the study of
the system of field equations
(\ref{eulerlagrangemetric}), (\ref{eulerlagrangeconnection}).

Consider the polarized Maxwell equation
\begin{equation}
\label{polarized Maxwell equation}
*\rmd A=\pm\rmi\rmd A
\end{equation}
in a classical pp-space, see Section \ref{Classical pp-waves}.
Here $A$ is the unknown complex vector field.
We seek plane wave solutions of (\ref{polarized Maxwell equation}),
see Definition \ref{definition of a plane wave}.
These can be written down explicitly:
\begin{equation}
\label{plane wave}
A=h(\varphi)\,a\,+\,k(\varphi)\,l\,.
\end{equation}
Here $l$ and $a$ are the vector fields defined in
Section \ref{Classical pp-waves}, $\ h,k:\mathbb{R}\to\mathbb{C}$ are arbitrary
functions, and $\varphi$ is the phase (\ref{phase}).

\begin{definition}
\label{definition of a generalised pp-space}
A \,\emph{generalised pp-space} is a metric compatible spacetime with
pp-metric and torsion
\begin{equation}
\label{define torsion}
T:=\frac12\mathrm{Re}(A\otimes\rmd A)
\end{equation}
where $A$ is a vector field of the form (\ref{plane wave}).
\end{definition}

We list below the main properties of generalised pp-spaces.
Here and further on we denote by $\{\!\nabla\!\}$ the covariant
derivative with respect to the Levi-Civita connection
which should not be confused with the full covariant derivative $\nabla$
incorporating torsion.

In the beginning of Section \ref{Classical pp-waves}
we introduced the spinor field $\chi$ satisfying $\{\!\nabla\!\}\chi=0$.
It turns out that this spinor field also satisfies $\nabla\chi=0$.
In other words, the generalised pp-space and underlying
classical pp-space admit the same nonvanishing parallel spinor field.
Consequently, both admit the same
nonvanishing parallel real null vector field $l$
and the same nonvanishing parallel complex 2-form $l\wedge a$.

The torsion of a generalised pp-space is purely tensor, i.e.
\begin{equation}
\label{purely tensor}
T^\alpha{}_{\alpha\gamma}=0,
\qquad
\varepsilon_{\alpha\beta\gamma\delta}T^{\alpha\beta\gamma}=0.
\end{equation}

The curvature of a generalised pp-space is
\begin{equation}
\label{curvature of a generalised pp-space}
R=-\frac12(l\wedge\{\!\nabla\!\})\otimes(l\wedge\{\!\nabla\!\})f
+\frac14\mathrm{Re}
\left(
(h^2)''\,(l\wedge a)\otimes(l\wedge a)
\right).
\end{equation}
Examination of formula (\ref{curvature of a generalised pp-space})
reveals the following remarkable properties of
generalised pp-spaces.
\begin{itemize}
\item
The curvatures generated by the Levi-Civita connection and torsion
simply add up
(compare formulae (\ref{curvature of a pp-space}) and
(\ref{curvature of a generalised pp-space})).
\item
The second term in the RHS of
(\ref{curvature of a generalised pp-space}) is purely Weyl.
Consequently, the Ricci curvature of a generalised pp-space is completely
determined by the pp-metric.
\item
The curvature of a generalised pp-space
has all the usual symmetries of curvature in the Riemannian case
(see Definition \ref{definition of riemannian}), that is,
\begin{eqnarray}
\label{symmetries of Riemannian curvature 1}
R_{\kappa\lambda\mu\nu}=R_{\mu\nu\kappa\lambda},
\\
\label{symmetries of Riemannian curvature 2}
\varepsilon^{\kappa\lambda\mu\nu}R_{\kappa\lambda\mu\nu}=0,
\\
\label{symmetries of Riemannian curvature 3}
R_{\kappa\lambda\mu\nu}=-R_{\lambda\kappa\mu\nu},
\\
\label{symmetries of Riemannian curvature 4}
R_{\kappa\lambda\mu\nu}=-R_{\kappa\lambda\nu\mu}.
\end{eqnarray}
Of course,
(\ref{symmetries of Riemannian curvature 4}) is true for any curvature
whereas (\ref{symmetries of Riemannian curvature 3})
is a consequence of metric compatibility.
Also, (\ref{symmetries of Riemannian curvature 3})
follows from
(\ref{symmetries of Riemannian curvature 1}) and
(\ref{symmetries of Riemannian curvature 4}).
\item
The second term in the RHS of (\ref{plane wave}) is pure gauge in the sense
that it does not affect curvature
(\ref{curvature of a generalised pp-space}).
It does, however, affect torsion (\ref{define torsion}).
\item
The Ricci curvature of a generalised pp-space is zero if and only if
\begin{equation}
\label{ricci is zero}
f_{11}+f_{22}=0
\end{equation}
and the Weyl curvature is zero if and only if
\begin{equation}
\label{weyl is zero}
f_{11}-f_{22}=\mathrm{Re}\left((h^2)''\right),
\qquad
f_{12}=-\frac12\mathrm{Im}\left((h^2)''\right).
\end{equation}
Here we use special local coordinates
(\ref{metric of a pp-wave}), (\ref{explicit l and a})
and denote
$f_{\alpha\beta}:=\partial_\alpha\partial_\beta f$.
\item
The curvature of a generalised pp-space is zero if and only if
we have both (\ref{ricci is zero}) and (\ref{weyl is zero}).
Clearly, for any given function $h$ we can choose a function $f$ such that $R=0$:
this $f$ is a quadratic polynomial in $x^1$, $x^2$ with coefficients
depending on $x^3$. Thus, as a spin-off, we get a class
of examples of Weitzenb\"ock spaces ($T\ne0$, $R=0$).
\end{itemize}

\section{Explicit representation of our field equations}
\label{Explicit representation of our field equations}

We write down explicitly our field equations
(\ref{eulerlagrangemetric}), (\ref{eulerlagrangeconnection})
under the following assumptions.
\begin{itemize}
\item
Our spacetime is metric compatible.
\item
Torsion is purely tensor, see (\ref{purely tensor}).
\item
Curvature has symmetries
(\ref{symmetries of Riemannian curvature 1}), (\ref{symmetries of Riemannian curvature 2}).
\item
Scalar curvature is zero.
\end{itemize}
Note that a generalised pp-space automatically possesses these properties.

It turns out that under the above assumptions the field equations are
\begin{eqnarray}
\label{eulerlagrangemetricLCexplicit}
\fl
d_1\mathcal{W}^{\kappa\lambda\mu\nu}Ric_{\kappa\mu}
+d_3\left(
Ric^{\lambda\kappa}Ric_\kappa{}^\nu
-\frac14g^{\lambda\nu}
Ric_{\kappa\mu}Ric^{\kappa\mu}\right)=0,
\\
\label{eulerlagrangeconnectionLCexplicit}
\fl
d_6\nabla_{\!\lambda}Ric_{\kappa\mu}
-d_7\nabla_{\!\kappa}Ric_{\lambda\mu}
\nonumber
\\
\lo
+d_6\left(
Ric^\eta{}_\kappa(T_{\eta\mu\lambda}-T_{\lambda\mu\eta})
+\frac12g_{\mu\lambda} \mathcal{W}^{\eta\zeta}{}_{\kappa\xi}
\left(T_\eta{}^\xi{}_\zeta-T_\zeta{}^\xi{}_\eta\right)
+g_{\mu\lambda}Ric^\eta{}_\xi T_\eta{}^\xi{}_\kappa
\right)
\nonumber
\\
\lo
-d_7\left(
Ric^\eta{}_\lambda(T_{\eta\mu\kappa}-T_{\kappa\mu\eta})
+\frac12g_{\kappa\mu}\mathcal{W}^{\eta\zeta}{}_{\lambda\xi}
\left(T_\eta{}^\xi{}_\zeta-T_\zeta{}^\xi{}_\eta\right)
+g_{\kappa\mu}Ric^\eta{}_\xi T_\eta{}^\xi{}_\lambda
\right)
\nonumber
\\
\lo
+ b_{10}\left(
\left(
g_{\kappa\mu}\mathcal{W}^{\eta\zeta}{}_{\lambda\xi}
-g_{\mu\lambda} \mathcal{W}^{\eta\zeta}{}_{\kappa\xi}
\right)
\left(T_\eta{}^\xi{}_\zeta-T_\zeta{}^\xi{}_\eta\right)
+Ric^\eta{}_\xi
\left(g_{\kappa\mu}T_\eta{}^\xi{}_\lambda-g_{\mu\lambda} T_\eta{}^\xi{}_\kappa\right)
\right)
\nonumber
\\
\lo
+
2b_{10}\left(
\mathcal{W}^\eta{}_{\mu\kappa\xi}
\left(T_\eta{}^\xi{}_\lambda - T_\lambda{}^\xi{}_\eta\right)
+\mathcal{W}^\eta{}_{\mu\lambda\xi}
\left(T_\kappa{}^\xi{}_\eta-T_\eta{}^\xi{}_\kappa\right)
-\mathcal{W}^{\xi\eta}{}_{\kappa\lambda}T_{\eta\mu\xi}
\right)=0,
\end{eqnarray}
where
\[
\begin{array}{ll}
d_1=b_{912}-b_{922}+b_{10},&\qquad d_3=b_{922}-b_{911},\\
d_6=b_{912}-b_{911}+b_{10},&\qquad d_7=b_{912}-b_{922}+b_{10},
\end{array}
\]
the $b$'s being coefficients from formula (51) of \cite{annalen}.
The LHS's of equations (\ref{eulerlagrangemetricLCexplicit}) and
(\ref{eulerlagrangeconnectionLCexplicit}) are the components of tensors $A$
and $B$ from the formula
\[
\delta S=\int(2A^{\lambda\nu}\,\delta g_{\lambda\nu}
+2B^{\kappa\mu}{}_\lambda\,\delta\Gamma^\lambda{}_{\mu\kappa})\,.
\]
Here $\delta g$ and $\delta \Gamma$ are the (independent) variations of the metric
and the connection, and $\delta S$ is the resulting variation of the action. In
(\ref{eulerlagrangeconnectionLCexplicit}) the first two indices of $B$
have been lowered to make the expression easier to read.

Equation (\ref{eulerlagrangemetricLCexplicit}) is
equation (12) of \cite{annalen} but with $\mathcal{R}=0$.
This is not surprising because when we vary the metric
it does not matter whether the curvature tensor
$R^\kappa{}_{\lambda\mu\nu}$ was generated by a Levi-Civita
connection or a general affine connection. What matters
are the symmetries
(\ref{symmetries of Riemannian curvature 1}),
(\ref{symmetries of Riemannian curvature 2})
which in our case
are the same as in the Riemannian case. In fact, our case
is simpler because scalar curvature is zero.

Equation (\ref{eulerlagrangeconnectionLCexplicit}) is similar to
equation (13) of \cite{annalen} but is not exactly the same.
Namely,
\begin{itemize}
\item
the first line of (\ref{eulerlagrangeconnectionLCexplicit}) coincides with
the LHS of equation (13) of \cite{annalen} with $\mathcal{R}\equiv0$,
\item
the remaining lines of (\ref{eulerlagrangeconnectionLCexplicit})
contain extra algebraic terms generated by torsion.
\end{itemize}
Note also that the covariant derivatives in
(\ref{eulerlagrangeconnectionLCexplicit}) and in
equation (13) of \cite{annalen} are different:
we use the notation $\nabla$ for the \emph{full} covariant derivative,
so the $\nabla$ in (\ref{eulerlagrangeconnectionLCexplicit})
itself incorporates torsion.
The arguments which produce
(\ref{eulerlagrangeconnectionLCexplicit})
are outlined in \ref{Derivation of the second field equation}.

\section{PP-wave type solutions of our field equations}
\label{PP-wave type solutions of our field equations}

The main result of this paper is the following

\begin{theorem}
\label{main theorem}
Generalised pp-spaces of parallel Ricci curvature are solutions of
the system of field equations
(\ref{eulerlagrangemetric}), (\ref{eulerlagrangeconnection}).
\end{theorem}

\noindent
{\it Proof\ }
The theorem is proved by direct substitution of
formulae for torsion, Ricci curvature and Weyl curvature
of a generalised pp-space
into the field equations
(\ref{eulerlagrangemetricLCexplicit}),~(\ref{eulerlagrangeconnectionLCexplicit}).
The $\nabla Ric$ terms in the LHS of
(\ref{eulerlagrangeconnectionLCexplicit})
vanish as Ricci curvature is assumed to be parallel,
so it remains to check the vanishing of
the remaining purely algebraic
terms in the LHS's of
(\ref{eulerlagrangemetricLCexplicit}),
(\ref{eulerlagrangeconnectionLCexplicit}).

According to Section \ref{PP-waves with torsion}
torsion, Ricci curvature and Weyl curvature
of a generalised pp-space are of the form
\begin{equation}
\label{torsion of a generalised pp-space}
T=\sum_{j,k=1}^2t_{jk}\,a_j\otimes(l\wedge a_k)
+\sum_{j=1}^2t_j\,l\otimes(l\wedge a_j),
\end{equation}
\begin{equation}
\label{Ricci curvature of a generalised pp-space}
Ric=s\,l\otimes l,
\end{equation}
\begin{equation}
\label{Weyl curvature of a generalised pp-space}
\mathcal{W}=\sum_{j,k=1}^2w_{jk}\,(l\wedge a_j)\otimes(l\wedge a_k),
\end{equation}
where $t_{jk}$, $t_j$, $s$, $w_{jk}$ are some real scalars satisfying
\[
t_{jk}=t_{kj},\qquad
w_{jk}=w_{kj},\qquad
t_{11}+t_{22}=w_{11}+w_{22}=0,
\]
$l$ and $a$ are vectors introduced in Section \ref{Classical pp-waves},
and $a_1=\mathrm{Re}\,a$, $a_2=\mathrm{Im}\,a$.
Note that the real vectors $l$, $a_1$, $a_2$ satisfy
\[
l\cdot l=l\cdot a_1=l\cdot a_2=a_1\cdot a_2=0,
\qquad
a_1\cdot a_1=a_2\cdot a_2=-1.
\]
All the algebraic terms containing $Ric$ in the LHS's of
(\ref{eulerlagrangemetricLCexplicit}),
(\ref{eulerlagrangeconnectionLCexplicit})
vanish because they involve contractions with
at least one of the indices of $Ric$, the latter
being of the form (\ref{Ricci curvature of a generalised pp-space})
with vector $l$ orthogonal to all other vectors
appearing in
(\ref{torsion of a generalised pp-space})--(\ref{Weyl curvature of a generalised pp-space}).
It remains to consider the $\mathcal{W}\times T$
terms in the LHS of
(\ref{eulerlagrangeconnectionLCexplicit}).
The terms with 3 contractions vanish
because in view of
(\ref{torsion of a generalised pp-space})
at least one of the contractions involves the vector $l$.
The term
$\mathcal{W}^{\xi\eta}{}_{\kappa\lambda}T_{\eta\mu\xi}$
also vanishes because in view of
(\ref{Weyl curvature of a generalised pp-space})
at least one of the contractions involves the vector $l$.
Thus, the proof of Theorem \ref{main theorem}
reduces to checking that
\begin{equation}
\label{reduces to checking that}
\mathcal{W}^\eta{}_{\mu\kappa\xi}
\left(T_\eta{}^\xi{}_\lambda - T_\lambda{}^\xi{}_\eta\right)
+\mathcal{W}^\eta{}_{\mu\lambda\xi}
\left(T_\kappa{}^\xi{}_\eta-T_\eta{}^\xi{}_\kappa\right)=0.
\end{equation}
The tensor in the LHS of
(\ref{reduces to checking that}) is proportional
to $l_\lambda l_\mu l_\kappa$ and is antisymmetric in
$\kappa$, $\lambda$, hence it is zero.
$\square$

Let $\{\!Ric\!\}$ denote the Ricci curvature
generated by the Levi-Civita connection
and let $\{\!\nabla\!\}$ denote, as usual, the covariant
derivative with respect to the Levi-Civita connection.
We know (see list of properties at the end of Section
\ref{PP-waves with torsion})
that in a generalised pp-space
$Ric=\{\!Ric\!\}$.
Moreover, it is easy to see that in a generalised pp-space
$\nabla Ric=\{\!\nabla\!\}Ric$.
This means that
when using Theorem \ref{main theorem} it does not
really matter whether the condition ``parallel Ricci curvature'' is understood
in the non-Riemannian sense $\nabla Ric=0$,
the Riemannian sense $\{\!\nabla\!\}\{\!Ric\!\}=0$,
or any combination of the two
($\{\!\nabla\!\}Ric=0$ or $\nabla\{\!Ric\!\}=0$).
In special local coordinates
(\ref{metric of a pp-wave}), (\ref{explicit l and a})
the condition that Ricci curvature is parallel is written as
$f_{11}+f_{22}=\mathrm{const}$ (compare with (\ref{ricci is zero})).

\section{Discussion}
\label{Discussion}

\subsection{Interpretation of our solutions}

Our interest in pp-spaces, classical and generalised, stems from our previous
publication~\cite{annalen}. It contained a comprehensive study of Riemannian
(see Definition \ref{definition of riemannian})
solutions of the field equations
(\ref{eulerlagrangemetric}), (\ref{eulerlagrangeconnection}).
It was shown in \cite{annalen} that the following
two classes of Riemannian spacetimes are solutions:
\begin{itemize}
\item
Einstein spaces ($Ric=\Lambda g$), and
\item
classical pp-spaces of parallel Ricci curvature.
\end{itemize}
Moreover, it was shown in \cite{annalen} that for a generic quadratic
action the above two classes of spacetimes are the \emph{only} Riemannian
solutions.

In General Relativity Einstein spaces are an accepted mathematical
model for vacuum. However, classical pp-spaces of parallel Ricci
curvature do not have an obvious physical interpretation.
Our current paper is an attempt at understanding whether
such spacetimes are of mathematical or physical significance.

Our analysis of vacuum solutions of quadratic metric-affine gravity
shows (Theorem~\ref{main theorem})
that classical pp-spaces of parallel Ricci curvature should not be
viewed on their own. They are a particular (degenerate) representative
of a wider class of solutions, namely,
generalised pp-spaces of parallel Ricci curvature.
The latter appear to admit a sensible physical interpretation.
Indeed, according to formula (\ref{curvature of a generalised pp-space})
the curvature of a generalised pp-space is a sum of two curvatures:
the curvature
\begin{equation}
\label{curvature of the underlying classical pp-space}
-\frac12(l\wedge\{\!\nabla\!\})\otimes(l\wedge\{\!\nabla\!\})f
\end{equation}
of the underlying classical pp-space
and the curvature
\begin{equation}
\label{curvature generated by a torsion wave}
\frac14\mathrm{Re}
\left(
(h^2)''\,(l\wedge a)\otimes(l\wedge a)
\right)
\end{equation}
generated by a torsion wave travelling over this classical pp-space.
Our torsion (\ref{define torsion}),~(\ref{plane wave})
and corresponding curvature (\ref{curvature generated by a torsion wave})
are waves travelling at speed of light because $h$~and~$k$ are functions of the
phase $\varphi$ which plays the role of a null coordinate,
$g^{\mu\nu}\nabla_\mu\varphi\,\nabla_\nu\varphi=0$, see formula (\ref{phase}).
The underlying classical pp-space of parallel Ricci curvature can now be
viewed as the ``gravitational imprint'' created by a wave of some
massless matter field.
Such a situation occurs in
Einstein--Maxwell theory
(classical model describing the interaction of gravitational and electromagnetic
fields)
and Einstein--Weyl theory
(classical model describing the interaction of gravitational and neutrino
fields).
The difference with our model is that
Einstein--Maxwell and Einstein--Weyl theories contain the gravitational
constant which dictates a particular relationship between the
strengths of the fields in question,
whereas our model is conformally invariant and the amplitudes of the
two curvatures
(\ref{curvature of the underlying classical pp-space})~and~(\ref{curvature generated by a torsion wave})
are totally independent.

The fundamental question is whether
torsion is a matter field, and, if it is, \emph{which} matter field.
In the remainder of this subsection we outline a
(highly speculative) argument in favour of interpreting
our torsion wave (\ref{define torsion}), (\ref{plane wave})
as a mathematical model for a neutrino field.

We base our interpretation on the analysis of the curvature
(\ref{curvature generated by a torsion wave})
generated by our torsion wave.
Examination of formula (\ref{curvature generated by a torsion wave})
indicates that it is more convenient to deal with the complexified
curvature
\begin{equation}
\label{complexified curvature generated by a torsion wave}
\mathfrak{R}:=r\,(l\wedge a)\otimes(l\wedge a)
\end{equation}
where $r:=\frac14(h^2)''$ (this $r$ is a function of
the phase $\varphi$);
note also that complexification
is in line with the traditions of quantum mechanics.
Our complex curvature is polarized,
\begin{equation}
\label{complex curvature is polarized}
{}^*\mathfrak{R}=\mathfrak{R}^*=\pm\rmi\mathfrak{R}\,,
\end{equation}
and purely Weyl, hence it is equivalent to a (symmetric) rank 4 spinor~$\omega$.
The relationship between $\mathfrak{R}$ and $\omega$ is
given by the formula
\begin{equation}
\label{spinor representation of curvature}
\mathfrak{R}_{\alpha\beta\gamma\delta}
=\sigma_{\alpha\beta ab}\,\omega^{abcd}\,\sigma_{\gamma\delta cd}
\end{equation}
where the
$\sigma_{\alpha\beta}$ are ``second order Pauli matrices''
(\ref{second order Pauli matrices}).
Resolving (\ref{spinor representation of curvature})
with respect to~$\omega$ we get,
in view of (\ref{complexified curvature generated by a torsion wave}),
(\ref{decomposition of F}), (\ref{formula for F}),
\begin{equation}
\label{formula for omega}
\omega=\xi\otimes\xi\otimes\xi\otimes\xi
\end{equation}
where
\begin{equation}
\label{formula for xi}
\xi:=r^{1/4}\,\chi
\end{equation}
and $\chi$ is the spinor field introduced
in the beginning of Section \ref{Classical pp-waves}.

Formula (\ref{formula for omega}) shows that our rank 4 spinor $\omega$
has additional algebraic structure:
it is the 4th tensor power of a rank 1 spinor $\xi$.
Consequently, the complexified curvature generated by our torsion
wave is completely determined by the rank 1 spinor field $\xi$.

We claim that the spinor field (\ref{formula for xi})
satisfies Weyl's equation, see
(\ref{Weyl equation 1}) or (\ref{Weyl equation 2}).
Indeed, as $\chi$ is parallel checking that $\xi$ satisfies Weyl's equation
reduces to checking that
$\,(r^{1/4})'\,\sigma^\mu{}_{a\dot b}\,l_\mu\,\chi^a=0\,$.
The latter is established by direct substitution of the
explicit formula for $l$, see (\ref{formula for l}).

\subsection{Comparison with existing literature}

There are a number of publications in which authors suggested various
generalisations of the concept of a classical pp-space. These generalisations
were performed within the Riemannian setting
(see Definition \ref{definition of riemannian})
and usually involved the incorporation of a constant nonzero scalar curvature;
see \cite{obukhov pp} and extensive further references therein.
Our construction described in Section \ref{PP-waves with torsion}
generalises the concept of a classical pp-space in a different
direction: we add torsion while retaining zero scalar curvature.

A powerful method which in the past has been used for the construction
of vacuum solutions of quadratic metric-affine gravity is the so-called
\emph{double duality ansatz}
\cite{mielkepseudoparticle,baekleretal1,baekleretal2,pseudo,annalen,mielkeduality}.
Its basic version \cite{pseudo} is as follows.
For certain types of quadratic actions
(see item (b) below)
the following is known to be true:
if the spacetime is metric compatible and curvature is irreducible
(i.e. all irreducible pieces except one are identically zero) then this
spacetime is a solution of
(\ref{eulerlagrangemetric}), (\ref{eulerlagrangeconnection}).
This fact is referred to as the double duality ansatz
because the proof is based on the use of the double duality
transform $R\mapsto{}^*\!R^*$ (this idea is due
to Mielke \cite{mielkepseudoparticle}) and because the
above conditions imply ${}^*\!R^*=\pm R$.
However, solutions presented
in Theorem~\ref{main theorem} do not fit into the double duality scheme.
This is due to the following reasons.
\begin{itemize}
\item[(a)]
The curvature of a pp-space, classical or generalised,
contains trace-free Ricci and Weyl pieces,
hence this curvature
is not necessarily irreducible and
not necessarily an eigenvector of the double duality operator.
Namely, for a pp-space the following statements are equivalent:
\begin{eqnarray*}
\!\!\!\!\!\!\!\!\!\!\!\!\!\!\!\!\!\!\!\!
\mbox
{$R$ is purely trace-free Ricci}\ &\Leftrightarrow\
\mbox
{condition (\ref{weyl is zero}) is satisfied}\ \Leftrightarrow\
\mbox{${}^*\!R^*=+R$}\,,
\\
\!\!\!\!\!\!\!\!\!\!\!\!\!\!\!\!\!\!\!\!
\qquad\mbox
{$R$ is purely Weyl}\ &\Leftrightarrow\
\mbox
{condition (\ref{ricci is zero}) is satisfied}\ \Leftrightarrow\
\mbox{${}^*\!R^*=-R$}\,.
\end{eqnarray*}
Furthermore, the curvature of a pp-space, classical or generalised,
does not necessarily satisfy the conditions of the modified double duality ansatz
\cite{baekleretal1,baekleretal2,mielkeduality}.
\item[(b)]
The double duality ansatz
in its basic \cite{pseudo} or modified \cite{baekleretal1,baekleretal2,mielkeduality}
forms does not work for the most general 16-parameter
actions introduced in \cite{Esser,hehlandmaciasexactsolutions2,annalen}
and considered in our current paper.
It works only for more special actions
with up to 11 coupling constants.
The fundamental difference between the 11-parameter and 16-parameter models
is best seen if one considers the specialisation of the
field equation (\ref{eulerlagrangeconnection}) to the Levi-Civita connection:
\begin{equation}
\label{eulerlagrangeconnectionLC}
\left.\partial S/\partial\Gamma\,\right|_{\mathrm{L-C}}=0.
\end{equation}
Equation (\ref{eulerlagrangeconnectionLC}) arises when one looks for Riemannian
solutions of (\ref{eulerlagrangeconnection}).
Here it is important to understand the logical sequence involved in
the derivation of (\ref{eulerlagrangeconnectionLC}):
we set
${\Gamma^\lambda}_{\mu\nu}=
\left\{{{\lambda}\atop{\mu\nu}}\right\}\,$
\emph{after} the variation of the connection has been carried out.
It is known \cite{pseudo} that for a generic 11-parameter action equation
(\ref{eulerlagrangeconnectionLC}) reduces to
\begin{equation}
\label{eulerlagrangeconnectionLC11}
\nabla_\lambda Ric_{\kappa\mu}-\nabla_\kappa Ric_{\lambda\mu}=0,
\end{equation}
whereas according to \cite{annalen} for a generic 16-parameter action equation
(\ref{eulerlagrangeconnectionLC}) reduces to
\begin{equation}
\label{eulerlagrangeconnectionLC16}
\nabla Ric=0.
\end{equation}
The field equations
(\ref{eulerlagrangeconnectionLC11})
and
(\ref{eulerlagrangeconnectionLC16})
are very much different, with (\ref{eulerlagrangeconnectionLC16})
being by far more restrictive.
In particular,
Nordstr\"om--Thompson spacetimes (Riemannian spacetimes with ${}^*\!R^*=+R$)
satisfy (\ref{eulerlagrangeconnectionLC11}) but do not necessarily
satisfy (\ref{eulerlagrangeconnectionLC16}).
\item[(c)]
The basic double duality ansatz \cite{pseudo} can be reformulated
in a way that makes it applicable to 16-parameter actions:
one has to impose the additional condition that curvature is \emph{simple},
i.e. the given irreducible subspace of the vector space of
curvatures is not isomorphic to any
other irreducible subspace.
See Section~6 of~\cite{annalen} for details.
According to formula (44) of~\cite{annalen} the (symmetric)
trace-free Ricci piece of curvature
is not simple,
hence the version of the double duality ansatz from \cite{annalen}
works for a pp-space, classical or generalised, only when
curvature is purely Weyl.
\end{itemize}

The new vacuum solutions of quadratic metric-affine gravity presented
in Theorem~\ref{main theorem} are similar to those of
Singh and Griffiths \cite{griffiths3}. The main differences are as follows.
\begin{itemize}
\item
The solutions in \cite{griffiths3} satisfy the condition
$\{\!Ric\!\}=0$ whereas our solutions satisfy the weaker
condition $\{\!\nabla\!\}\{\!Ric\!\}=0$
(see also last paragraph
of Section \ref{PP-wave type solutions of our field equations}).
\item
The solutions in \cite{griffiths3} were obtained
for the Yang--Mills case (\ref{YMq}) whereas we deal with
a general $\mathrm{O}(1,3)$-invariant quadratic form $q$
with 16 coupling constants.
\end{itemize}

The observation that one can construct vacuum solutions of
quadratic metric-affine gravity in terms of pp-waves is
a recent development. The fact that classical pp-spaces of parallel
Ricci curvature are solutions was first pointed out in
\cite{garda,poland,annalen}.

\ack

The authors are grateful to  J B Griffiths and F W Hehl for helpful advice.

\appendix

\section{Spinor formalism for generalised pp-spaces}
\label{Spinor formalism for generalised pp-spaces}

In this appendix, unless otherwise stated,
we work in a general metric compatible spacetime
with torsion. We start by recalling basic facts about spinors.

Define the ``metric spinor''
\begin{equation}
\label{metric spinor}
\epsilon_{ab}=\epsilon_{\dot a\dot b}=
\epsilon^{ab}=\epsilon^{\dot a\dot b}=
\left(
\begin{array}{cc}
0&1\\
-1&0
\end{array}
\right)
\end{equation}
with the first index enumerating rows and the second enumerating columns.
We raise and lower spinor indices according to the formulae
\begin{equation}
\label{raising and lowering of spinor indices}
\xi^a=\epsilon^{ab}\xi_b,
\qquad
\xi_a=\epsilon_{ab}\xi^b,
\qquad
\eta^{\dot a}=\epsilon^{\dot a\dot b}\eta_{\dot b},
\qquad
\eta_{\dot a}=\epsilon_{\dot a\dot b}\eta^{\dot b}.
\end{equation}
Our definition
(\ref{metric spinor}), (\ref{raising and lowering of spinor indices})
has the following advantages.
\begin{itemize}
\item
The spinor inner product is invariant under the operation of raising and
lowering of indices, i.e.
$(\epsilon_{ac}\xi^c)(\epsilon^{bd}\eta_d)=\xi^a\eta_b$.
\item
The ``contravariant'' and ``covariant'' metric spinors are
``raised'' and ``lowered'' versions of each other, i.e.
$\epsilon^{ab}=\epsilon^{ac}\epsilon_{cd}\epsilon^{bd}$
and
$\epsilon_{ab}=\epsilon_{ac}\epsilon^{cd}\epsilon_{bd}$.
\end{itemize}
The disadvantage of our definition
(\ref{metric spinor}), (\ref{raising and lowering of spinor indices})
is that the consecutive raising and lowering of a single spinor index leads
to a change of sign, i.e. $\epsilon_{ab}\epsilon^{bc}\xi_c=-\xi_a$.
This inconsistency is related to the well known fact that a spinor does
not have a particular sign (say, a spatial rotation of the coordinate
system by $2\pi$ leads to a change of sign). In formulae where the sign
is important we will be careful in specifying our choice of sign; see,
for example, (\ref{definition of contravariant Pauli matrices}),
(\ref{spinor connection coefficient}).

Let $\mathfrak{v}$ be the real vector space of Hermitian $2\times2$
matrices $\sigma_{a\dot b}$. Pauli matrices
$\sigma^\alpha{}_{a\dot b}$, $\alpha=0,1,2,3$,
are a basis in $\mathfrak{v}$ satisfying
$\sigma^\alpha{}_{a\dot b}\sigma^{\beta c\dot b}
+\sigma^\beta{}_{a\dot b}\sigma^{\alpha c\dot b}
=2g^{\alpha\beta}\delta_a{}^c$
where
\begin{equation}
\label{definition of contravariant Pauli matrices}
\sigma^{\alpha a\dot b}:=
\epsilon^{ac}\sigma^\alpha{}_{c\dot d}\epsilon^{\dot b\dot d}.
\end{equation}
At every point of the manifold $M$ Pauli matrices are defined uniquely up to
a Lorentz transformation.
For the pp-metric (\ref{metric of a pp-wave}) we choose Pauli matrices
\begin{equation}
\label{Pauli matrices for pp-metric}
\fl
\sigma^0{}_{a\dot b}=
\left(
\begin{array}{cc}
1&0\\
0&-f
\end{array}
\right),
\
\sigma^1{}_{a\dot b}=
\left(
\begin{array}{cc}
0&1\\
1&0
\end{array}
\right),
\
\sigma^2{}_{a\dot b}=
\left(
\begin{array}{cc}
0&\mp\rmi\\
\pm\rmi&0
\end{array}
\right),
\
\sigma^3{}_{a\dot b}=
\left(
\begin{array}{cc}
0&0\\
0&2
\end{array}
\right).
\end{equation}
Our two choices of Pauli matrices differ by orientation. When dealing with
a classical pp-space the choice of orientation of Pauli matrices
does not really matter, however in a generalised pp-space it is convenient
to choose orientation of Pauli matrices in agreement with the signs in
(\ref{polarized Maxwell equation})
and (\ref{complex curvature is polarized})
as this simplifies the resulting formulae.

Define
\begin{equation}
\label{second order Pauli matrices}
\sigma_{\alpha\beta ac}:=\frac12
\bigl(
\sigma_{\alpha a\dot b}\epsilon^{\dot b\dot d}\sigma_{\beta c\dot d}
-
\sigma_{\beta a\dot b}\epsilon^{\dot b\dot d}\sigma_{\alpha c\dot d}
\bigr)\,.
\end{equation}
These ``second order Pauli matrices'' are polarized, i.e.
\begin{equation}
\label{polarization of second order Pauli matrices}
*\sigma=\pm i\sigma
\end{equation}
depending on the orientation of ``basic'' Pauli matrices
$\sigma^\alpha{}_{a\dot b}$, $\alpha=0,1,2,3$.
Note that with our choice of Pauli matrices
the signs in formulae (\ref{Pauli matrices for pp-metric})
and (\ref{polarization of second order Pauli matrices}) agree.

We define the covariant derivatives of spinor fields as
\[
\nabla_\mu\xi^a=\partial_\mu\xi^a+\Gamma^a{}_{\mu b}\xi^b,
\qquad
\nabla_\mu\xi_a=\partial_\mu\xi_a-\Gamma^b{}_{\mu a}\xi_b,
\]
\[
\nabla_\mu\eta^{\dot a}=\partial_\mu\eta^{\dot a}
+\bar\Gamma^{\dot a}{}_{\mu\dot b}\eta^{\dot b},
\qquad
\nabla_\mu\eta_{\dot a}=\partial_\mu\eta_{\dot a}
-\bar\Gamma^{\dot b}{}_{\mu\dot a}\eta_{\dot b},
\]
where
$\bar\Gamma^{\dot a}{}_{\mu\dot b}=\overline{\Gamma^a{}_{\mu b}}$.
The explicit formula for the spinor connection coefficients
$\Gamma^a{}_{\mu b}$ can be derived from the following two conditions:
\begin{equation}
\label{condition 1}
\nabla_\mu\epsilon^{ab}=0,
\end{equation}
\begin{equation}
\label{condition 2}
\nabla_\mu j^\alpha=\sigma^\alpha{}_{a\dot b}\nabla_\mu\zeta^{a\dot b},
\end{equation}
where $\zeta$ is an arbitrary rank 2 mixed spinor field and
$j^\alpha:=\sigma^\alpha{}_{a\dot b}\zeta^{a\dot b}$
is the corresponding vector field (current).
Conditions (\ref{condition 1}), (\ref{condition 2})
give a system of linear algebraic equations for
$\mathrm{Re}\,\Gamma^a{}_{\mu b}$, $\mathrm{Im}\,\Gamma^a{}_{\mu b}$
the unique solution of which is
\begin{equation}
\label{spinor connection coefficient}
\Gamma^a{}_{\mu b}=\frac14
\sigma_\alpha{}^{a\dot c}
\left(
\partial_\mu\sigma^\alpha{}_{b\dot c}
+\Gamma^\alpha{}_{\mu\beta}\sigma^{\beta}{}_{b\dot c}
\right).
\end{equation}
In a generalised pp-space formula (\ref{spinor connection coefficient}) reads
as follows: the nonzero coefficients are
\[
\Gamma^1{}_{12}=\frac12hh',
\qquad
\Gamma^1{}_{22}=\mp \frac\rmi 2hh',
\qquad
\Gamma^1{}_{32}=
\frac12\left(\frac{\partial f}{\partial x^1}
\pm\rmi\frac{\partial f}{\partial x^2}\right)
-\frac12kh'.
\]
Here we use special local coordinates
(\ref{metric of a pp-wave}), (\ref{explicit l and a})
and Pauli matrices (\ref{Pauli matrices for pp-metric}).

The generally accepted point of view
\cite{
hehl habilitation,
hehl neutrino 1,
hehl neutrino 2,
hehl neutrino 3,
griffiths neutrino}
is that a neutrino field
in a metric compatible spacetime with or without torsion
is described by the action
\[
S_\mathrm{neutrino}:=2i\int
\Bigl(
\xi^a\,\sigma^\mu{}_{a\dot b}\,(\nabla_\mu\bar\xi^{\dot b})
\ -\
(\nabla_\mu\xi^a)\,\sigma^\mu{}_{a\dot b}\,\bar\xi^{\dot b}
\Bigr),
\]
see formula (11) of \cite{griffiths neutrino}.
Variation in $\xi$ produces Weyl's equation
\[
\sigma^\mu{}_{a\dot b}\nabla_\mu\,\xi^a
-\frac12T^\eta{}_{\eta\mu}\sigma^\mu{}_{a\dot b}\,\xi^a=0
\]
which can be equivalently rewritten as
\[
\sigma^\mu{}_{a\dot b}\{\!\nabla\!\}_\mu\,\xi^a
\pm\frac\rmi 4
\varepsilon_{\alpha\beta\gamma\delta}T^{\alpha\beta\gamma}
\sigma^\delta{}_{a\dot b}\,\xi^a=0
\]
where $\{\!\nabla\!\}$ is the covariant derivative with respect to the Levi-Civita
connection. In a generalised pp-space torsion is purely tensor,
see (\ref{purely tensor}),
so Weyl's equation takes the form
\begin{equation}
\label{Weyl equation 1}
\sigma^\mu{}_{a\dot b}\nabla_\mu\,\xi^a=0
\end{equation}
or, equivalently,
\begin{equation}
\label{Weyl equation 2}
\sigma^\mu{}_{a\dot b}\{\!\nabla\!\}_\mu\,\xi^a=0.
\end{equation}

\section{Derivation of the second field equation}
\label{Derivation of the second field equation}

In this Appendix we outline the arguments which produce
(\ref{eulerlagrangeconnectionLCexplicit}).
Throughout this Appendix the metric is assumed to be fixed
and the connection is being varied.
We also assume that we start variation from a spacetime
satisfying the four conditions listed in the beginning of
Section \ref{Explicit representation of our field equations}.

Following the reasoning of Section 3 of \cite{annalen},
we rewrite our quadratic form as
\begin{equation}
\label{formula for quadratic form with dots}
\!\!\!\!\!\!\!\!\!\!\!\!\!
q(\!R)
=c_1(R^{(1)},R^{(1)})_{\mathrm{YM}}
+c_3(R^{(3)},R^{(3)})_{\mathrm{YM}}
+2(b_{911}-b_{922})(P_-,P_+)+\ldots
\end{equation}
where $(\,\cdot\,,\,\cdot\,)_{\mathrm{YM}}$ is the Yang--Mills inner product on curvatures
$(R,Q)_{\mathrm{YM}}:=R^\kappa{}_{\lambda\mu\nu}\,
Q^\lambda{}_\kappa{}^{\mu\nu}\,$
and $\ldots$ denote terms which do not contribute to $\delta S$.
Here the $R^{(j)}$s are the irreducible pieces of curvature
labelled in accordance with \cite{pseudo}.
The tensors $P_\pm$ are defined by
$P_-:=\frac12(\mathcal{R}ic-\mathcal{R}ic^{(2)})$,
$P_+:=\frac12(\mathcal{R}ic+\mathcal{R}ic^{(2)})
=\frac12(Ric+Ric^{(2)}),$
where
$Ric^{(2)}{}_{\kappa\nu}:=R_\kappa{}^\lambda{}_{\lambda\nu}$,
$\mathcal{R}ic^{(2)}:=Ric^{(2)}+\frac14\mathcal{R}g$,
and the constants $c_1$, $c_3$ are given by
\begin{equation}
\label{formluae for cs}
c_1=-\frac12(b_{911}-2b_{912}+b_{922}),
\qquad
c_3=b_{10}
\end{equation}
in agreement with formula (15) of \cite{annalen}.
The $b$'s are coefficients from formula (51) of \cite{annalen}.

The variations of $\int(R^{(j)},R^{(j)})_{\mathrm{YM}}$
were computed in Section~4
of \cite{pseudo}:
\begin{equation}
\label{variation of terms from the pseudoinstanton paper}
\delta\int(R^{(j)},R^{(j)})_{\mathrm{YM}}
=4\int\tr\,\bigl(
(\delta_{\mathrm{YM}}R^{(j)})^\mu\,(\delta\Gamma)_\mu
\bigr)
\end{equation}
where
$(\delta_\mathrm{YM}R)^\mu:=\frac1{\sqrt{|{\det}g|}}\,
(\partial_\nu+[\Gamma_\nu\,,\,\cdot\,])
(\sqrt{|\det g|}\,R^{\mu\nu})$
is the Yang--Mills divergence.
Here we hide the Lie algebra indices of curvature by using matrix notation;
say, $[\Gamma_\xi\,,R_{\mu\nu}]$ stands for
\begin{equation}
\label{hiding indices}
[\Gamma_\xi\,,R_{\mu\nu}]^\kappa{}_\lambda=
\Gamma^\kappa{}_{\xi\eta}R^\eta{}_{\lambda\mu\nu}
-R^\kappa{}_{\eta\mu\nu}\Gamma^\eta{}_{\xi\lambda}.
\end{equation}
Now, in our case
${R^{(1)}}_{\kappa\lambda\mu\nu}=\frac12
(g_{\kappa\mu}Ric_{\lambda\nu}
-g_{\lambda\mu}Ric_{\kappa\nu}
-g_{\kappa\nu}Ric_{\lambda\mu}
+g_{\lambda\nu}Ric_{\kappa\mu})$,
$R^{(3)}=\mathcal{W}$,
with the other $R^{(j)}$'s being zero.
Substituting these expressions into
(\ref{variation of terms from the pseudoinstanton paper})
we get
\begin{eqnarray}
\fl
\delta\int(R^{(1)},R^{(1)})_{\mathrm{YM}}=
2 \int
(\nabla_\lambda Ric_{\kappa\mu}
- \nabla_\kappa Ric_{\lambda\mu}
+ g_{\kappa\mu} \nabla_\eta Ric_\lambda{}^\eta
- g_{\lambda\mu} \nabla_\eta Ric_\kappa{}^\eta
\nonumber
\\
\label{variation 1}
\quad \ \
+ Ric_\kappa{}^\eta ( T_{\eta\mu\lambda} - T_{\lambda\mu\eta})
+ Ric_\lambda{}^\eta (T_{\kappa\mu\eta} - T_{\eta\mu\kappa} ))
\ \delta \Gamma^{\lambda\mu\kappa},
\end{eqnarray}
\begin{equation}
\label{variation 2}
\fl
\delta\int(R^{(3)},R^{(3)})_{\mathrm{YM}}
=4\int
(\nabla_\eta \mathcal{W}_{\kappa\lambda\mu}{}^\eta
+\mathcal{W}_{\kappa\lambda}{}^{\eta\xi}T_{\eta\mu\xi})
\ \delta \Gamma^{\lambda\mu\kappa}.
\end{equation}
The variation of $\int(P_-,P_+)$ turns out to be
\begin{eqnarray}
\label{variation 3}
\delta\int(P_-,P_+)=\int(Ric,\delta P_+)
=\frac12\int(Ric,\delta Ric)+\frac12\int(Ric,\delta Ric^{(2)})
\nonumber
\\
=
-\frac12\int
[\nabla_\lambda Ric_{\kappa\mu}
+\nabla_\kappa Ric_{\lambda\mu}
-g_{\mu\kappa}\nabla_\eta Ric_\lambda{}^\eta
-g_{\mu\lambda}\nabla_\eta Ric_\kappa{}^\eta
\nonumber
\\
\qquad\qquad\qquad
+Ric_\kappa{}^\eta(T_{\eta\mu\lambda} - T_{\lambda\mu\eta})
+Ric_\lambda{}^\eta(T_{\eta\mu\kappa}-T_{\kappa\mu\eta})
]\ \delta\Gamma^{\lambda\mu\kappa}
\end{eqnarray}
(compare with the corresponding formula in
Section 3 of \cite{annalen}).

Combining formulae
(\ref{formula for quadratic form with dots}),
(\ref{formluae for cs}),
(\ref{variation 1})--(\ref{variation 3}) we arrive at
the explicit form of the field equation
(\ref{eulerlagrangeconnection}):
\begin{eqnarray}
\label{eulerlagrangeconnectionLCexplicitoriginal}
\fl
d_6' (\nabla_\lambda Ric_{\kappa\mu} - g_{\mu\lambda} \nabla_\eta Ric^\eta{}_\kappa
-T_{\lambda\mu\eta}Ric^\eta{}_\kappa + T_{\eta\mu\lambda}Ric^\eta{}_\kappa)
\nonumber
\\
\lo
-d_7' (\nabla_\kappa Ric_{\lambda\mu} - g_{\kappa\mu}\nabla_\eta Ric^\eta{}_\lambda
-T_{\kappa\mu\eta}Ric^\eta{}_\lambda + T_{\eta\mu\kappa}Ric^\eta{}_\lambda)
\nonumber
\\
\lo
+2b_{10} (\nabla_\eta \mathcal{W}^\eta{}_{\mu\lambda\kappa} -
\mathcal{W}^{\eta\xi}{}_{\kappa\lambda} T_{\xi\mu\eta})=0
\end{eqnarray}
where
\[
d_6' = b_{912}-b_{911},\qquad d_7' = b_{912}-b_{922}.
\]

Let us now make use of the Bianchi identity for curvature
\begin{equation}
\label{bianchi}
(\partial_\xi+[\Gamma_\xi\,,\,\cdot\,])R_{\mu\nu}
+
(\partial_\nu+[\Gamma_\nu\,,\,\cdot\,])R_{\xi\mu}
+
(\partial_\mu+[\Gamma_\mu\,,\,\cdot\,])R_{\nu\xi}
=0
\end{equation}
where we hide the Lie algebra indices of curvature by using matrix notation
as in (\ref{hiding indices}).
Making one contraction in (\ref{bianchi})
and using the four assumptions listed in the beginning of
Section \ref{Explicit representation of our field equations}
we get
\begin{eqnarray}
\label{bianchi1}
\fl
\frac12[ \nabla_\kappa Ric_{\mu\lambda}
-\nabla_\lambda Ric_{\mu\kappa}
+g_{\mu\kappa}\nabla_\eta Ric^\eta{}_\lambda
-g_{\mu\lambda}\nabla_\eta Ric^\eta{}_\kappa
\nonumber
\\
\lo
+Ric^\eta{}_\xi (g_{\mu\kappa}T_\eta{}^\xi{}_\lambda
-g_{\mu\lambda}T_\eta{}^\xi{}_\kappa)
+Ric^\eta{}_\kappa(T_{\eta\lambda\mu}-T_{\lambda\eta\mu})
+Ric^\eta{}_\lambda(T_{\kappa\eta\mu} -T_{\eta\kappa\mu})]
\nonumber
\\
\lo
+\nabla_\eta \mathcal{W}^\eta{}_{\mu\lambda\kappa}
+\mathcal{W}^\eta{}_{\mu\kappa\xi}(T_\lambda{}^\xi{}_\eta - T_\eta{}^\xi{}_\lambda)
+ \mathcal{W}^\eta{}_{\mu\lambda\xi}(T_\eta{}^\xi{}_\kappa -T_\kappa{}^\xi{}_\eta)=0.
\end{eqnarray}
Another contraction in (\ref{bianchi1}) yields
\begin{equation}
\label{covariantRicContraction}
\nabla_\eta Ric^\eta{}_\lambda =
-Ric^\eta{}_\xi T_\eta{}^\xi{}_\lambda
- \frac12 \mathcal{W}^{\eta\zeta}{}_{\lambda\xi}
(T_\eta{}^\xi{}_\zeta - T_\zeta{}^\xi{}_\eta).
\end{equation}
Substitution of (\ref{covariantRicContraction}) into (\ref{bianchi1}) gives
\begin{eqnarray}
\label{bianchi2}
\fl
\nabla_\eta \mathcal{W}^\eta{}_{\mu\lambda\kappa} =
\mathcal{W}^\eta{}_{\mu\kappa\xi}(T_\eta{}^\xi{}_\lambda -T_\lambda{}^\xi{}_\eta)
+\mathcal{W}^\eta{}_{\mu\lambda\xi}(T_\kappa{}^\xi{}_\eta-T_\eta{}^\xi{}_\kappa)
\nonumber
\\
\lo
+\frac14 (T_\zeta{}^\xi{}_\eta -T_\eta{}^\xi{}_\zeta)
(g_{\mu\lambda} \mathcal{W}^{\eta\zeta}{}_{\kappa\xi}
-g_{\mu\kappa} \mathcal{W}^{\eta\zeta}{}_{\lambda\xi})
\nonumber
\\
\lo
+\frac12[\nabla_\lambda Ric_{\mu\kappa}
-\nabla_\kappa Ric_{\mu\lambda}
+Ric^\eta{}_\kappa(T_{\lambda\eta\mu}-T_{\eta\lambda\mu})
+Ric^\eta{}_\lambda(T_{\eta\kappa\mu}-T_{\kappa\eta\mu})].
\end{eqnarray}

Formulae (\ref{covariantRicContraction}) and
(\ref{bianchi2}) allow us to exclude the
terms with
$\nabla_\eta Ric^\eta{}_\kappa$,
$\nabla_\eta Ric^\eta{}_\lambda$
and $\nabla_\eta \mathcal{W}^\eta{}_{\mu\lambda\kappa}$
from equation
(\ref{eulerlagrangeconnectionLCexplicitoriginal})
reducing the latter to (\ref{eulerlagrangeconnectionLCexplicit}).

\end{document}